\documentclass[prb,twocolumn,showpacs,preprintnumbers,amsmath,amssymb,floats,floatfix]{revtex4-1}
\pdfoutput=1
\usepackage{graphicx}
\usepackage{amsmath}
\usepackage{epstopdf}
\usepackage{amsbsy}
\usepackage[usenames,dvipsnames]{xcolor}
\usepackage{pstricks}
\usepackage{hyperref}
\usepackage{dcolumn}
\usepackage{bm}
\newcommand{\BFCA}{Ba(Fe$_{1-x}$Co$_x$)$_2$As$_2$}

\newcommand{\Q}{{\bf Q}}

\def\k{{\bf k}}

\def\q{{\bf q}}

 \graphicspath{{./29_06_2015/}}

\begin{document}

\title{Spin excitations in a model of FeSe with orbital ordering}
\author{A. Kreisel$^1$, Shantanu Mukherjee$^1$, P. J. Hirschfeld$^2$ and Brian M. Andersen$^1$}
\affiliation{
$^1$Niels Bohr Institute, University of Copenhagen, Juliane Maries Vej 30, DK-2100 Copenhagen, Denmark\\
$^2$Department of Physics, University of Florida, Gainesville, Florida 32611, USA
}

\date{January 2, 2016}

\begin{abstract}
We present a theoretical study of the dynamical spin susceptibility for the intriguing Fe-based superconductor FeSe, based on a tight-binding model developed to account for the temperature-dependent band structure in this system.
The model allows for orbital ordering in the $d_{xz}/d_{yz}$ channel below the structural transition and presents
a strongly $C_4$ symmetry broken Fermi surface at low temperatures which accounts for the nematic
properties of this material. The calculated spin excitations are peaked at wave vector $(\pi,0)$
in the 1-Fe Brillouin zone, with a broad maximum at energies of order a few meV. In this range,
the occurrence of superconductivity sharpens  this peak in energy, creating  a $(\pi,0)$ ``neutron resonance'' as seen in recent experiments.
With the exception of the quite low energy scale of these fluctuations, these results are roughly similar to standard behavior in Fe pnictide systems.  At higher
energies, however, intensity increases and shifts to wave vectors along the $(\pi,0)$ - $(\pi,\pi)$ line. We compare with existing inelastic neutron experiments and NMR data, and give predictions for further studies.
\end{abstract}


\maketitle
\section{Introduction}
The structurally simplest Fe-based superconductor, FeSe, is one of the most mysterious at the present writing.   It exhibits a tetragonal to orthorhombic structural phase transition at $T_S\sim 90\,\text{K}$, and displays very strong electronic nematic behavior below this temperature ($T$), but never orders magnetically as do the more familiar Fe pnictide systems.  While its critical temperature  $T_c\sim 9\,\text{K}$ is relatively low for this class of materials, a very rapid increase of  $T_c$ to about $40\,\text{K}$ is observed under modest pressure\cite{medvedev09}. Various intercalates of FeSe have $T_c$ of roughly this magnitude at ambient pressure as well, and the highest $T_c$ values of the entire class of Fe-based superconductors, $70\text{-}100\,\text{K}$,  are found in monolayer films of FeSe grown on SrTiO$_3$ substrates.\cite{yan12,tan13,ge14}  Thus the bulk FeSe material, of which excellent single crystals are now available\cite{bohmer13}, gives the impression of being poised to become a high-temperature superconductor, such that efforts to
understand its properties have accelerated in the past few years.

A starting point for these theoretical efforts is a reasonable band structure for FeSe. Unfortunately, both density functional theory (DFT) and dynamical mean field theory (LDA+DMFT) predict low-energy bands that deviate qualitatively from that observed by angle-resolved photoemission  (ARPES)\cite{nakayama14,maletz14,shimojima14,watson14,ZhangP15,SuzukiY15,ZhangY15} and quantum oscillation (QO) studies\cite{terashima14,watson14}. At high $T$, ARPES finds splittings of the hole band associated with Fe $d_{xz}/d_{yz}$ states that are consistent with
a modest spin-orbit coupling of approximately $20\,\text{meV}$. At low $T$, an additional splitting that breaks the fourfold $d_{xz}/d_{yz}$ symmetry is also observed at the M-point and it increases significantly and smoothly in a manner consistent with orbital ordering. The Fermi pocket sizes are generally much smaller than found in {\it ab initio} studies, and show a  $C_4$ symmetry breaking significantly larger than one might anticipate given the ${\mathit O}(0.3\%$) orthorhombic splitting of the lattice
constants.  In Ref. \onlinecite{MKHA_PRL15}, we presented a tight-binding band structure ``engineered'' to give the correct ARPES and QO results, both at high and low $T$, including a $T$-dependent orbital order term added to the Hamiltonian, and showed that the model gave reasonable results also for the Knight shift and spin-lattice relaxation rate $(T_1T)^{-1}$ compared to experiment\cite{baek14,bohmer15}. Using this band structure we also calculated the spin fluctuation exchange pairing interaction, and showed that it yielded a superconducting ground state with gap structure consistent with STM and penetration depth measurements.\cite{MKHA_PRL15}

Recently, the first inelastic neutron scattering (INS) results on this system became available\cite{Boothroyd15,Bourges15,Wang2015}, showing the dominance of ``conventional'' $(\pi,0)$ spin fluctuations at low energies and persisting up to $T_S$, while previous {\it ab initio} approaches did not find low-energy excitations at that momentum\cite{Essenberger12}. For $T$ greater than $T_S$, both INS and NMR experiments have suggested that no significant low-energy spin fluctuation weight was present, in contrast to pnictide systems such as BaFe$_2$As$_2$,
where such strong magnetic fluctuations were tied to the nematic transition\cite{Fernandes_etal_nematic14}. Understanding the cause for suppression of $(\pi,0)$ spin fluctuations seen in neutron scattering and NMR at high $T$ is clearly important to understand the physical reasons for the suppression of magnetic order
in these systems, and the relation, if any, between spin fluctuations, nematic order and orbital order. Several novel proposals have  been made recently about how unusual magnetic effects might prevent the occurrence of long range magnetic order\cite{Glasbrenner15,Kivelson15,Chubukov15} or lead to a type of order that is difficult to observe\cite{Si_quadrupolar15}.

The $(\pi,0)$ spin fluctuations in FeSe have a broad maximum, according to INS, at very low energy, of order $3-4\,\text{meV}$, which acquires
significant amplitude  immediately below $T_S$\cite{Bourges15}.   This is in apparent conflict with NMR spin relaxation results, which probe low-energy
spin fluctuations at all ${\bf q}$, but which are negligible until lower $T$ just above $T_c$ are reached\cite{baek14,bohmer15}.  One proposal to reconcile the results of these two probes is the presence of an apparent spin gap in the INS measurements below about 2 meV; there is no apparent
explanation for such a spin gap, however. Another possibility is  that the spin fluctuations remain at intermediate $T$ and NMR spin relaxation is less sensitive to them in this material due to the presence of  competing fluctuations that evolve differently with the evolution of $T$-dependent band structure. When the $T$ is lowered below $T_c$,  the $(\pi,0)$  fluctuations sharpen into a resonance similar to those observed in Fe pnictide systems. It is remarkable, in fact, that at low $T$ and low energies, the magnetic response of the system is similar to canonical iron-based systems such as \BFCA\cite{Inosov15_review,Dai15_review}

In this paper, we present results for the dynamical susceptibility  derived from our band model for FeSe together with a random phase approximation  (RPA) treatment of Hubbard and Hund-type interactions.  We reproduce the low-energy $(\pi,0)$ spin fluctuations and the neutron resonance in the superconducting state,  but show that at higher energies the spin fluctuations disperse to wave vectors along the $(\pi,0)-(\pi,\pi)$ line in momentum space.  We continue to find a $T$-dependence of the Knight shift and $(T_1T)^{-1}$ in agreement with experiment, despite the fact that our model does not have a low-energy normal state spin gap.  We attribute the apparent insensitivity of the NMR to the $(\pi,0)$ fluctuations at high $T$ to the weak overall weight at this wave vector, due to the electronic structure of this material.
The superconducting order parameter derived from spin-fluctuation pairing has $s+d$ character with accidental nodes, leading to
a V-shaped density of states (DOS) at the Fermi surface. Nodal lines occur and thus lead in our calculations to
anisotropies in the low-temperature penetration depth, varying as $T$ and $T^3$ in the $y$ and $x$ directions,
respectively.  These predictions can  be checked experimentally in a single-domain sample.
The overall picture that emerges for this material is a system driven to the verge of a magnetic instability by orbital ordering.

\section{Model}
The Hamiltonian for this system is given by
\begin{eqnarray}
H&=&H_{TB} + H_{OO} + H_U,
\end{eqnarray}
where $H_{TB}$ is the tight-binding Hamiltonian, $H_{OO}$ is a $T$-dependent orbital ordering term that breaks the crystal point-group symmetry, and $H_U$ is the standard Hubbard-Hund Hamiltonian, see Appendix A. The tight-binding term can be expressed as
\begin{eqnarray}
H_{TB}&=&\sum_{\bf{k},\mu,\nu,\sigma}t_{\mu\nu}({\bf{k}})c^{\dag}_{\mu\sigma}({\bf{k}})c_{\nu\sigma}({\bf{k}}).
\end{eqnarray}
Here ($\mu,\nu$) are orbital indices, $t_{\mu\nu}({\bf{k}})$ are the hopping integrals, and $n_{\mu\sigma}({\bf{k}})=c^{\dag}_{\mu\sigma}({\bf{k}})c_{\mu\sigma}({\bf{k}})$. A band structure that is consistent with the observed electronic structure in FeSe has been provided in Ref.~\onlinecite{MKHA_PRL15}, where it was shown that the rather complex renormalizations of band structure relative to
DFT observed in Fe-based superconducting systems\cite{BorisenkoSO}, involving both Fermi velocity changes and relative shifts of hole and electron bands, could be captured in terms of relatively simple modifications of a few short-range matrix elements (details can be found in Appendix A).
The model assumes further that below the structural transition at $T\sim 90\,\text{K}$ the fourfold-symmetry-broken phase is described by a mixed bond- and site-centered $T$-dependent orbital ordering term in the Hamiltonian given by
\begin{align}
 H_{OO}&=\Delta_{b}g(t)\sum_{{\bf{k}}\sigma}(\cos k_x-\cos k_y )(n_{xz\sigma}({\bf{k}})+n_{yz\sigma}({\bf{k}})) \nonumber\\
       & +\Delta_{s}g(t)\sum_{{\bf{k}}\sigma}(n_{xz\sigma}({\bf{k}})-n_{yz\sigma}({\bf{k}})).
       \label{eq_oo}
\end{align}
Here $\Delta_{b}$ ($\Delta_{s}$) is the bond-centered (site-centered) orbital ordering which we assume has a mean-field $T$ dependence $g(t)$ below $T_S=90\;\text{K}$ with $t=T/T_S$,\cite{Note1}
and leads to a maximum band splitting of $50\,\text{meV}$. This combination of orbital ordering is motivated by a recent ARPES measurement on detwinned FeSe\cite{SuzukiY15}, showing that the band splitting between the $d_{xz}/d_{yz}$ bands occurs in opposite directions at the $\Gamma$ and M points. We find a similar behavior in our calculated band energies $\xi_\nu(\bf{k})$  as discussed in Appendix A. Band splittings consistent with the results in Ref.~\onlinecite{SuzukiY15} are reproduced by $\Delta_s(T=0)=\Delta_b(T=0)=50\;\text{meV}$, together with a spin-orbit coupling of $20\;\text{meV}$ determined by the high-$T$ bands at the $\Gamma$ point.

\section{Results}

\subsection{Static spin susceptibility}
\begin{figure}[t]
\includegraphics[width=\columnwidth]{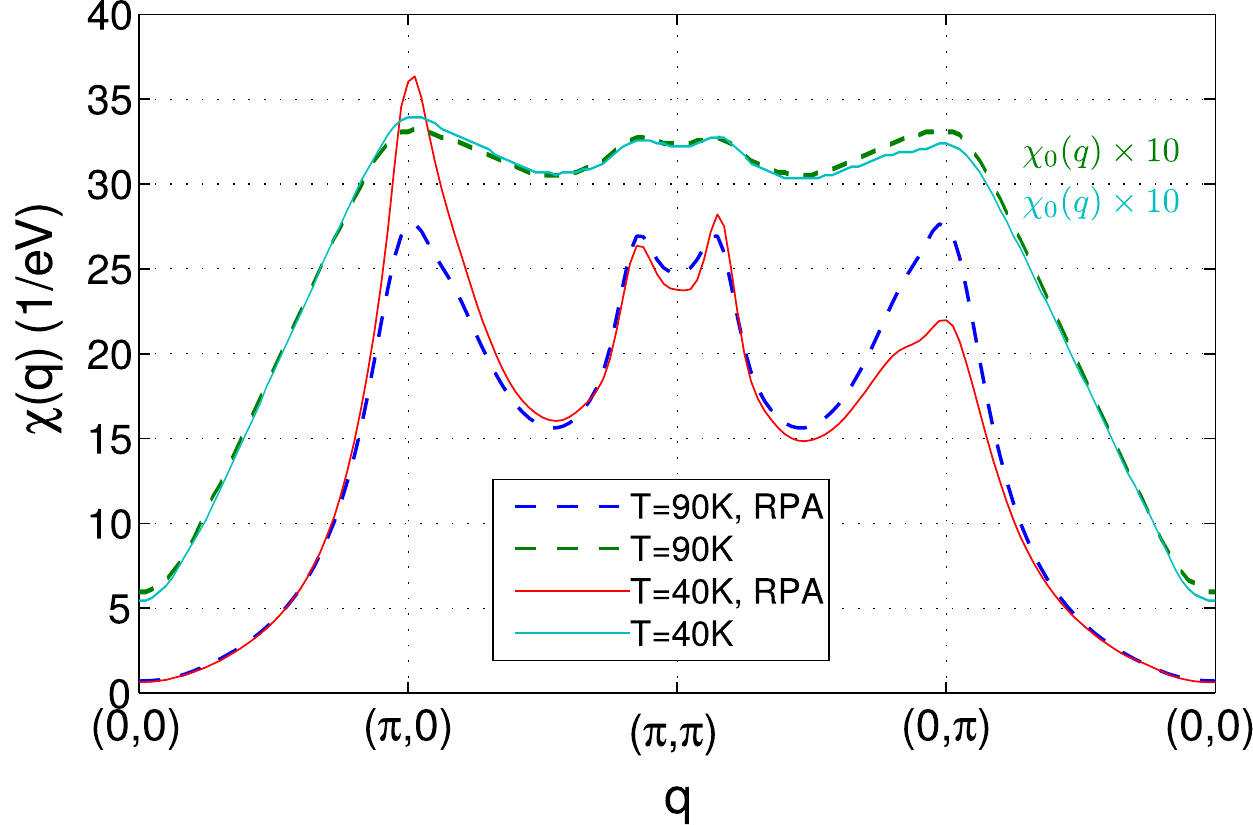}
\caption{(Color online) Real part of bare static susceptibilities $\chi_0(\q,0)$ and those in RPA approximation $\chi_{\text{RPA}}(\q,0)$ at $90\,\text{K}$ and $40\,\text{K}$. The bare susceptibilities have been multiplied by a factor of 10 to improve readability of the plot.}
\label{fig:staticChi}
\end{figure}
Focusing first on the static, $\omega=0$, magnetic properties, Fig.~\ref{fig:staticChi} shows
the bare and RPA spin susceptibilities $\chi_0(\q,0)$ and $\chi_{\text{RPA}}(\q,0)$ as a function of momentum. It is striking that in the high-$T$ ($C_4$ symmetric) magnetic response,
the $(\pi,0)$ instability competes very closely with a state near $(\pi,\pi)$ [$(\pi,q)$, where $q$ is close to $\pi$], even when interactions are included close to the RPA instability.    This
is reminiscent of the scenario of Glasbrenner \textit{et al.},\cite{Glasbrenner15} whereby FeSe is nonmagnetic due to competition of several nearly degenerate magnetic states
with wave vector $(\pi,q)$.  As $T$ is lowered, this degeneracy is lifted by orbital ordering, as evident from Fig.~\ref{fig:staticChi}, singling out the $(\pi,0)$ fluctuations as the dominant ones. This remains true also with the susceptibility enhanced by interactions in the RPA scheme; spectral weight is moved downwards in energy near $(\pi,0)$ to dominate the spin spectrum up to $100\;\text{meV}$ as seen in Fig.~\ref{fig:highneut}. From the Fermi surface plots shown in Fig. \ref{fig:bandfscut} in Appendix A it can be inferred that there is no obvious nesting at the Fermi surface that can readily explain the larger noninteracting susceptibility at $(\pi,0)$ compared to $(0,\pi)$. This property originates from scattering at higher energies. We will return to this point further below.

\subsection{Dynamical spin susceptibility: normal state}
\begin{figure}[tb]
\includegraphics[width=\columnwidth]{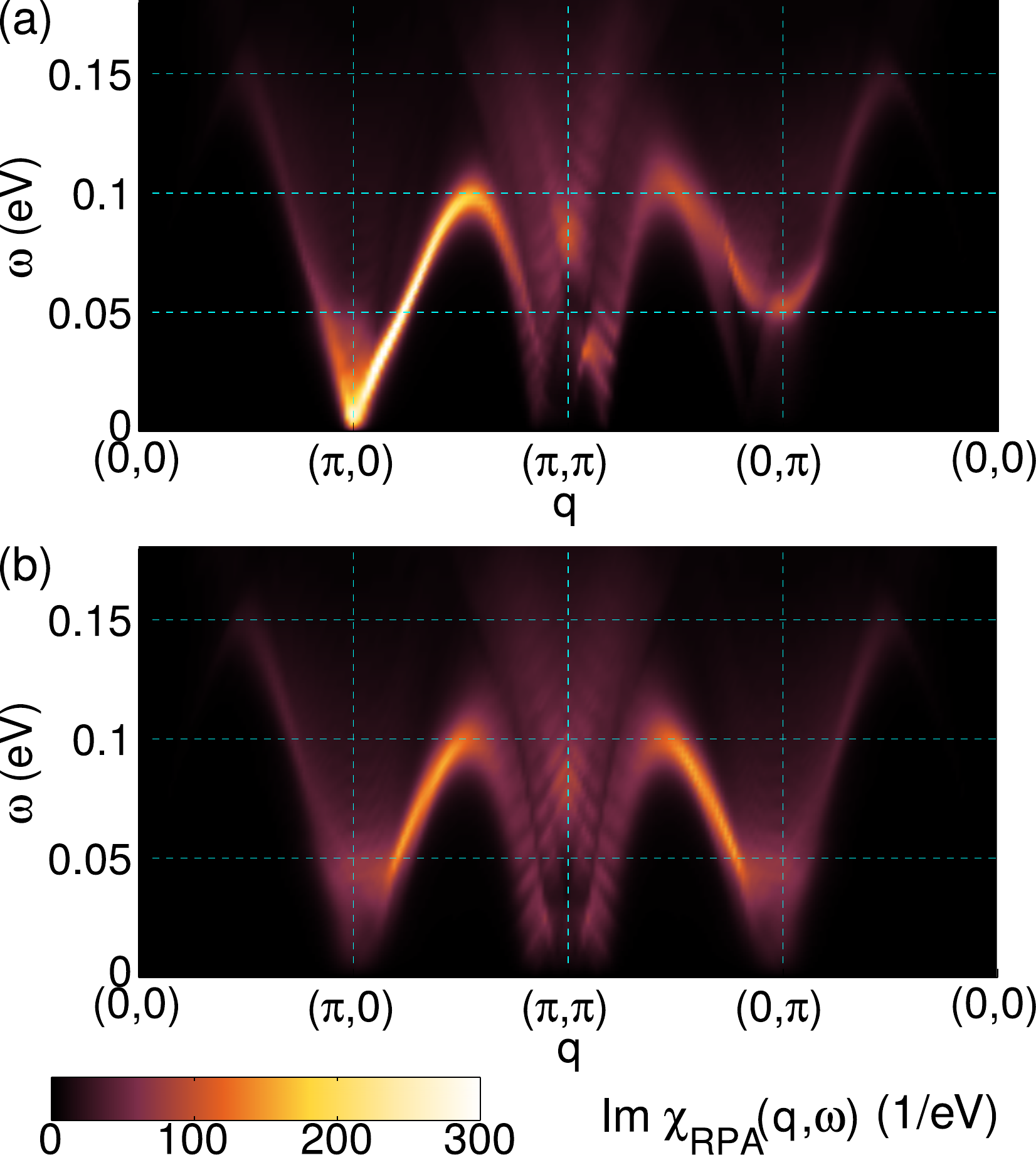}
\caption{(Color online) (a) Imaginary part of the paramagnetic susceptibility $\mathop{\text{Im}}\chi_{\text{RPA}}$ in $(\text{eV})^{-1}$ along high symmetry directions at $T=40\;\text{K}$.
(b) Susceptibility calculated at $T=110\;\text{K}$ above the structural transition.
}
\label{fig:highneut}
\end{figure}
Inelastic neutron scattering (INS) experiments on FeSe have revealed the importance of conventional $(\pi,0)$ fluctuations\cite{Boothroyd15,Bourges15,Wang2015}. Experiments performed on powder samples have found $(\pi,0)$ fluctuations extending to at least 80 meV. As can be seen in Fig.~\ref{fig:highneut} we also find similar $(\pi,0)$ fluctuations extending up to $\sim 100\,\text{meV}$ when the interactions are tuned close to the Stoner instability ($U=0.3515z\;\text{eV}$, $J=0.25U$, $z=6$ is the band renormalization factor) such that there is no magnetic instability at least down to the temperature of the superconducting transition. Note that at these high energies, the $(\pi,0)$ excitations
are present both below and above the structural transition, as in experiment\cite{Boothroyd15,Wang2015}. We will come back to differences in weights of the stripe fluctuations close to $(\pi,0)$ and N\'eel fluctuations close to $(\pi,\pi)$ as a function of $T$ later. Additionally, the imaginary part of the susceptibility disperses and becomes more intense along the line from $(\pi,0)$ to $(\pi,\pi)$ as we move to higher energies. Furthermore, there are weaker incommensurate branches which soften near $(\pi,\pi)$, in contrast to observed spin excitations in Fe pnictides\cite{Dai15_review}.
This feature of the high-energy excitations constitute a prediction of the present theory for INS experiments on single crystals extending to higher energies.

\begin{figure}[tb]
\includegraphics[width=\columnwidth]{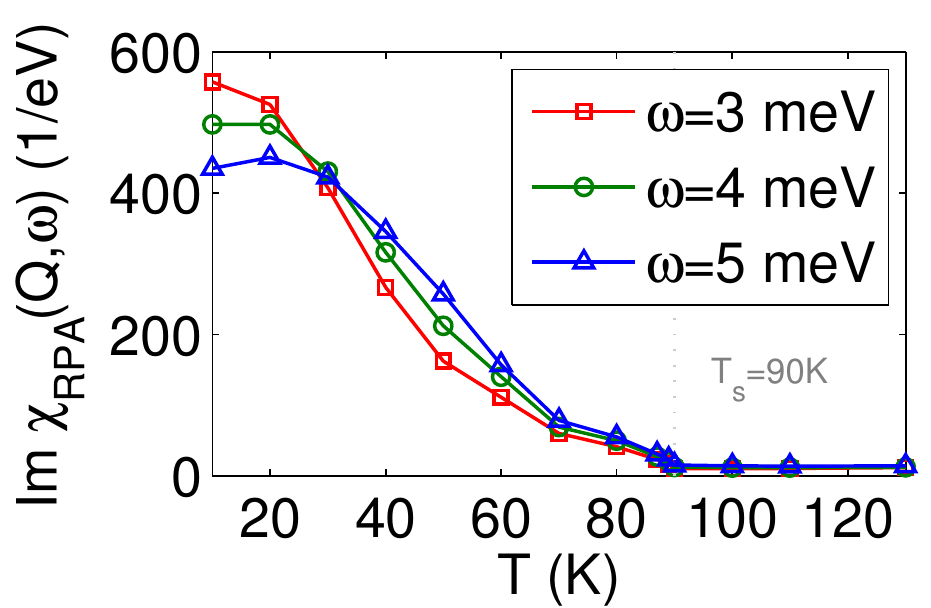}
\caption{(Color online)
$T$ dependence of the imaginary part of the  dynamical susceptibility  $\mathop{\text{Im}}\chi_{\text{RPA}}(\bf{Q},\omega)$ at $\Q=(\pi,0)$ calculated from our model with the interaction parameters as in Fig. \ref{fig:highneut}, shown for series
of low energies $\omega=3,4,5\,\text{meV}$.
 }
\label{fig:lowneut}
\end{figure}
It is interesting to observe that above the structural transition, the low-energy  $(\pi,0)$ fluctuations are strongly suppressed [see Fig.~\ref{fig:highneut}(b)], in agreement with experiments\cite{Bourges15}. This suppression can be more clearly seen in Fig.~\ref{fig:lowneut} where the $T$ dependence of the low-energy spin fluctuations at $(\pi,0)$ is plotted. The strength of the low-energy $(\pi,0)$ fluctuations dies off rapidly with increasing
$T$, and disappears completely above $T_S$.  
Note that the increase in $\mathop{\text{Im}}\chi_{\text{RPA}}$ at low $T$ with decreasing $\omega$ is due to the low-energy peak in $\mathop{\text{Im}}\chi_{\text{RPA}}$  since the system is close to the magnetic instability.

\setcounter{figure}{4}
\begin{figure*}[tb]
\includegraphics[width=2.08\columnwidth]{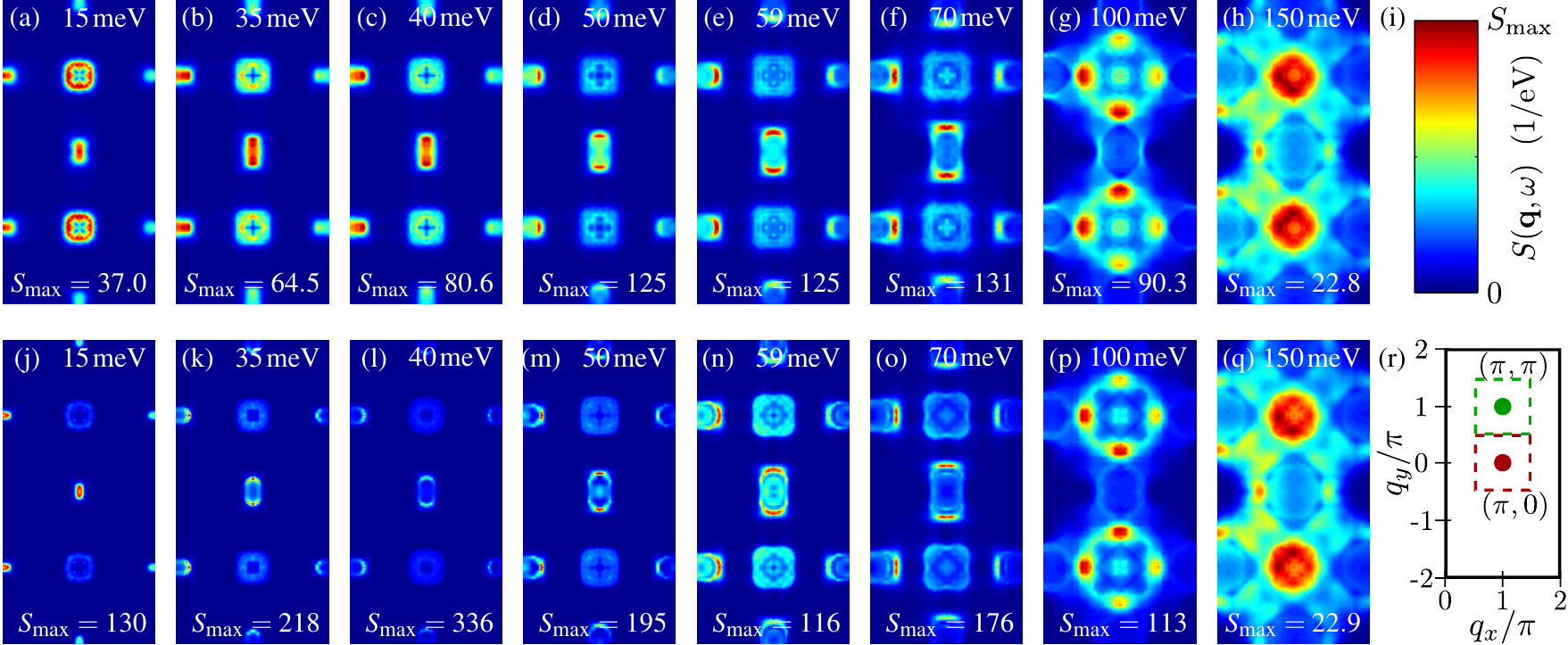}
\caption{(Color online) Dynamical structure factor  $S({\bf q},\omega)$ plotted as function of momentum $\bf q$ ($q_z=0$) for various energies $\omega$ as marked at the individual figures at $T=110\;\text{K}$ (a-h) and  $T=10\;\text{K}$ (j-q). All plots range over the same $q$ space as shown in (r) and have the same color scale (i) where the maximum $S_{\text{max}}$ is indicated at the bottom of each plot. To allow an easier comparison to experimental data as presented in Ref. \onlinecite{Wang2015}, the results for $T$ below the structural transition have been averaged over two domains of orbital order to simulate twinned crystals. The two regions around $(\pi,0)$ and $(\pi,\pi)$ indicated in (r) by dashed lines are used as integration areas to deduce the local stripe and N\'eel fluctuations.
\label{fig:S_maps}}
\end{figure*}

This unusual $T$ dependence is primarily due to an enhancement of low-energy spin fluctuations induced by the orbital ordering. By introducing correlations that push the system close to the magnetic instability, the low-$T$ fluctuations are amplified more strongly compared to  high $T$. We argue therefore that the suppression of the high-$T$ spin fluctuations seen in NMR experiments is not  necessarily due to the presence of a spin gap, but
may arise simply due to $T$-dependent electronic structure effects.   For example, we note that while the $(\pi,0)$ fluctuations
strengthen at low $T$, those near $(\pi,\pi)$ lose spectral weight.  Since both such fluctuations contribute to the NMR spin relaxation,
they evidently compensate each other to some extent, leaving a flatter $T$ dependence in the intermediate-$T$ region; thus $(T_1T)^{-1}$ does
not rise until close to $T_c$ when the $(\pi,0)$ fluctuations completely dominate, as discussed in Appendix B.

\setcounter{figure}{3}
\begin{figure}[t]
\includegraphics[width=\columnwidth]{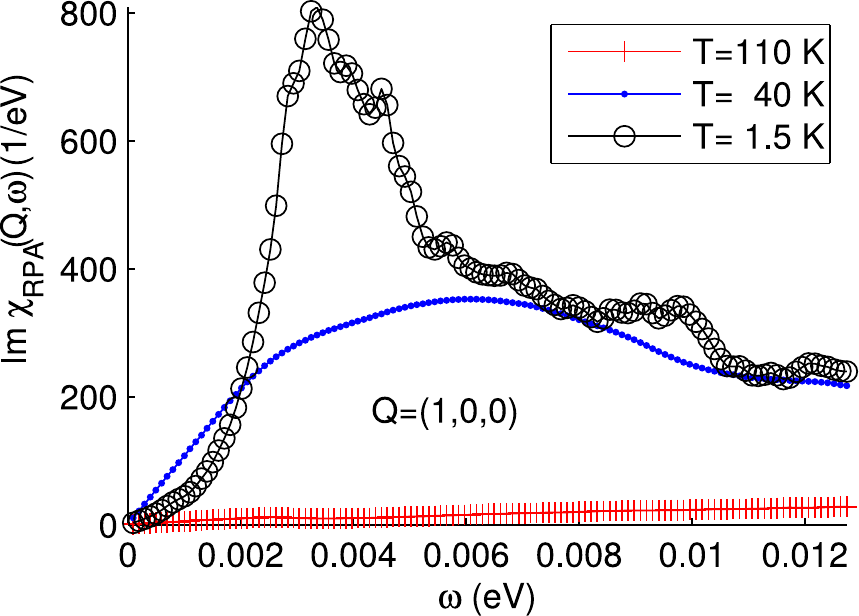}
\caption{(Color online)
 Spin fluctuation spectral weight  $\mathop{\text{Im}} \chi_{\text{RPA}}(\bf{Q}, \omega)$ at 
$\Q=(\pi,0)$ for the same interaction parameters as in Fig.~\ref{fig:highneut} at three representative values of $T$. For the lowest temperature ($T=1.5\;\text{K}$), $\mathop{\text{Im}} \chi_{\text{RPA}}(\bf{Q}, \omega)$ is calculated in the presence of the superconducting gap within the spin-fluctuation pairing approach as detailed in Ref. \onlinecite{Maier08}.
\label{fig:nresonance}}
\end{figure}

In Fig.~\ref{fig:nresonance} the energy dependence of the  low-energy  inelastic spectrum  is shown at $T=40\;\text{K}$. It displays a peak at around $\omega=4\;\text{meV}$,  similar to INS experiments\cite{Bourges15}. Note that such a peak occurs in our model primarily because the Coulomb interaction is tuned close to the Stoner instability. Therefore within such an itinerant model the presence of the low-energy peak structure is a signature that FeSe is very close to a magnetic instability,  consistent also with pressure experiments\cite{Cava09,Bendele12,Terashima15}. Recently, the magnetic excitations in FeSe have been studied in more detail via INS experiments.\cite{Wang2015}. For a closer comparison, we also present maps of the dynamical structure factor $S({\bf q},\omega)$ at different energies and two temperatures above and below the structural transition in \mbox{Fig. \ref{fig:S_maps}}. The dynamical structure factor has been calculated using
\setcounter{figure}{5}
\begin{figure}[t]
\includegraphics[width=\columnwidth]{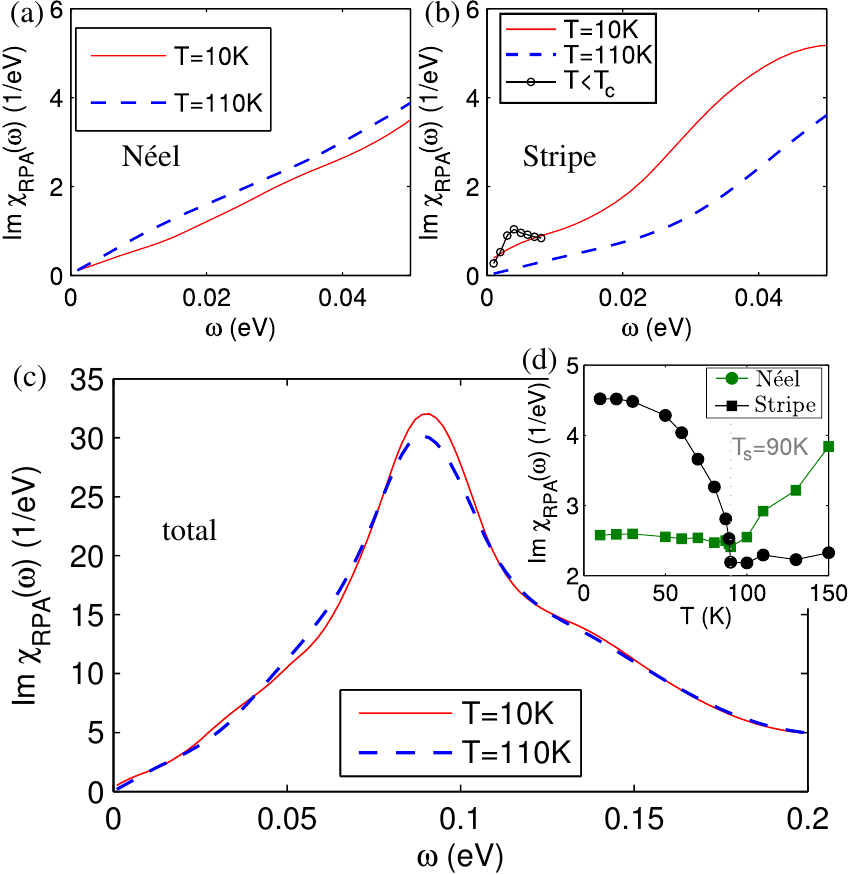}
\caption{(Color online)
Local susceptibility $\mathop{\text{Im}}\chi_{\text{RPA}}(\omega)$ at two representative temperatures above and below the structural transition obtained by integration over momentum space at $q_z=0$. In (a) we show the contributions of N\'eel fluctuations by integrating over one quarter of the Brillouin zone centered around $(\pi,\pi)$ compare Fig. \ref{fig:S_maps} (r). In (b) the area is centered around $(\pi,0)$ to estimate the stripe fluctuations, showing a transfer of spectral weight to stripe fluctuations due to orbital order, and a low-energy peak in the superconducting state. The total local susceptibility (c) is unaffected by lowering $T$, but the weight transfer as a function of $T$ can be observed clearly at an energy $\omega=40\;\text{meV}$ (d).
\label{fig:integrated_susc}}
\end{figure}
\begin{equation}
S({\bf q},\omega)= \frac{1}{1-e^{-\omega/T}}f^2({\bf q})\mathop{\text{Im}}\chi_{\text{RPA}}({\bf q},\omega)
\end{equation}
where $f({\bf q})$ is the magnetic form factor of $\text{Fe}^{2+}$, such that the result does not show periodicity with reciprocal vectors as the susceptibility does\cite{TabCrys}. In Fig. \ref{fig:S_maps}, we show plots of the dynamical structure factor as a function of momentum for various energies at high $T$ and low $T$. For low energies $\omega < 50\;\text{meV}$ it can be seen that the spot around $(\pi,0)$ is elongated perpendicular to the direction $(0,0)$-$(\pi,0)$ at all temperatures showing the dispersion of the fluctuations along the path $(\pi,0)$ to $(\pi,\pi)$, which can also be seen in Fig. \ref{fig:highneut}. Already from this series of plots one can see that the spectral weight around $(\pi,\pi)$ is present at all temperatures, but decreases when orbital order sets in and the fluctuations at $(\pi,0)$ sharpen and gain weight. To make this effect more visible, we also calculate the local susceptibility $\chi_{\text{RPA}}(\omega)=\int_{A} d^2{\bf q}\, \chi_{\text{RPA}}({\bf q},\omega)$, where the integration has been performed for $q_z=0$ since the susceptibility shows only a very weak dependence on $q_z$. In Fig. \ref{fig:integrated_susc} we present the results for partial integrals over square regions $A$ of one quarter of the Brillouin zone around $(\pi,\pi)$ to estimate the weight of the N\'eel fluctuations (a) and around $(\pi,0)$ to estimate the stripe fluctuations (b). The regions of integration are indicated in Fig. \ref{fig:S_maps} (r) with dashed lines.

Restricting ourselves to the energy range where the two types of fluctuations are separated in momentum space, one can observe that the  N\'eel fluctuations decrease in weight as orbital order sets in, while the stripe fluctuations gain weight. The total local susceptibility is nearly unchanged over the whole bandwidth of the paramagnon excitation spectrum that extends up to energies of approximately $200\;\text{meV}$ with a maximum at around  $90\;\text{meV}$; see  Fig. \ref{fig:integrated_susc} (c). In summary, our results agree extremely well with experimental results presented in Ref. \onlinecite{Wang2015} in the transfer of the spectral weight from  N\'eel fluctuations to stripe fluctuations on lowering $T$, see Fig. \ref{fig:integrated_susc} (d), while the former are already weaker above the structural transition. Also the total local susceptibility is in agreement with the experimental result. Note that the relative strength of N\'eel fluctuations and stripe fluctuations changes slightly with the ratio $J/U$ and details of the electronic structure such that quantitative conclusions cannot be drawn, but the trend of weight transfer is robust.

To conclude this discussion, we mention that our result slightly deviates from experimental findings in Ref. \onlinecite{Wang2015} in the nature of the N\'eel fluctuations where they seem to be gapped at low $T$ and do not deviate from a commensurate structure. We expect however that the incommensurate peaks near $(\pi,\pi)$ will coalesce in our model if self-energy effects from disorder and interactions are included in the model, but this calculation is beyond the scope of this paper.

\subsection{Dynamical spin susceptibility: superconducting state}

\begin{figure}[bt]
\includegraphics[width=\columnwidth]{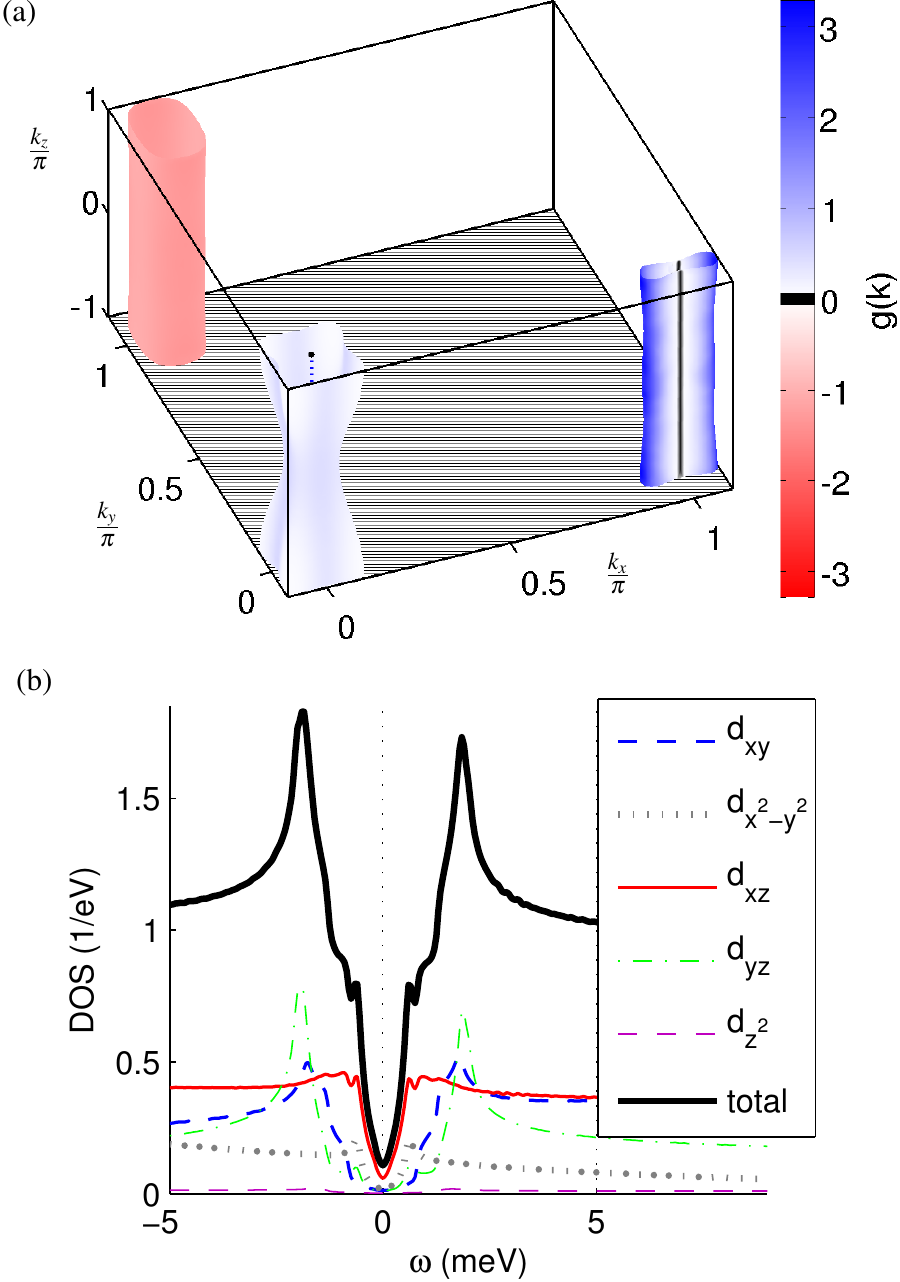}
\caption{(Color online) (a) Gap symmetry function on the Fermi surface $g({\bf k})$ of the leading superconducting instability with $\lambda=0.15$ as obtained from the solution of the linearized gap equation with $U=0.32z\;\text{eV}$ and $J=0.25U$, where $z=6$ is the band renormalization factor. Areas where $g({\bf k})$ is smaller than 1\% of the maximal value are marked black. (b) DOS calculated by solving the full BdG equation with the same interaction parameters in momentum space using a grid of $12\times 12\times 6$ $k$ points and interpolating the converged gap to calculate it with a reasonable spectral resolution.}
\label{fig_SC}
\end{figure}

Next, we calculate the superconducting order parameter using spin-fluctuation-mediated pairing interactions. Within our previous approach of solving the linearized gap equation
using a five-band model for computational simplicity\cite{MKHA_PRL15}, we obtain the gap symmetry function $g(\bf{k})$ on the Fermi surface as shown in Fig.~\ref{fig_SC}(a). Due to the reduction of the crystal symmetry by the orbital order parameter at low $T$, this state has components of both $s$- and $d$-wave symmetry. In our model, the superconducting gap is very small on the $\Gamma$-centered pocket and obeys line nodes at the $(\pi,0)$ pocket. The sign change between the electron-like bands between $(\pi,0)$ and $(0,\pi)$ is induced primarily by pair scattering originating from the broad $(\pi,\pi)$ structure of the susceptibility. From the larger value of the susceptibility at momentum transfer $(\pi,0)$ compared to $(0,\pi)$ (see Fig. \ref{fig:staticChi}) one would naively expect a sign change between the $\Gamma$-centered pocket and the pocket at $(\pi,0)$. Instead, the sign change appears between the hole-like band and electron-like band with Fermi surface centered at $(0,\pi)$ because of the effect of matrix elements in the pair-scattering vertex; i.e., it is driven by the orbital content at the Fermi level.

Next, we solve the Bogoliubov-de Gennes (BdG) equations in momentum space following the method in Ref.~\onlinecite{Romer2015} for the multiorbital case with symmetrized pairing interactions in order to calculate the DOS which shows a V-shaped spectral dependence at low energies as seen in Fig.~\ref{fig_SC}(b). This nodal V-shaped DOS is in agreement with recent STM measurements.\cite{song11,kasahara14}

For the calculation of the imaginary part of the susceptibility in the superconducting state, we use the framework discussed in Ref.~\onlinecite{Maier08}. The gap symmetry function from the linearized gap equation is interpolated on a large momentum mesh, and we employ an exponential damping factor\cite{Maier09} to specify its value away from the Fermi surface with damping chosen one order of magnitude larger than the gap $\Delta_0$. We chose a smearing $\eta\approx 1\,\text{K}$ and use an integration grid of $\approx 10^6$ $k$ points to obtain reasonable results as shown in Fig.~\ref{fig:nresonance}. At high $T$ above the structural transition, there is very little  low-energy spectral weight at $(\pi,0)$.  It  gradually increases when the $T$ is decreased and orbital order sets in, such that at lower $T$ a maximum at a few meV is obtained. Deep in the superconducting state where the superconducting gap is fixed to yield coherence peaks at $\approx 2.2\;\text{meV}$, see Fig.~\ref{fig_SC}(b), spectral weight at low energies is suppressed due to the superconducting gap opening and the maximum is again at approximately $4\;\text{meV}$, which also induces a low-energy peak in the local stripe fluctuations, see Fig. \ref{fig:integrated_susc} (b), but not in the local N\'eel fluctuations. The low-$T$ contributions near  $(\pi,\pi)$ (not shown) are very small, similar to the paramagnetic state shown in Figs.~\ref{fig:highneut} and \ref{fig:S_maps} (j).

To conclude our discussion of the low-energy spin fluctuations,
we emphasize that the increase of the peak upon lowering of $T$ into the superconducting state is somewhat different
from the usual neutron resonance expected in the Fe pnictides, where a mode corresponding to a pole in the RPA susceptibility is allowed
to propagate due to the gapping of the Fermi surface in the superconducting state, and the effect is weighted by a coherence factor $\sum_\k \left[1 - \frac{\Delta_\k \Delta_{\k+\Q}}{E_{\k} E_{\k+\Q}}\right]$ that is maximized when the gap changes sign between hole and electron pockets.  In the case of FeSe, at $\Q=(\pi,0)$ these coherence factors play a negligible role since the gap changes sign between the hole pocket and the electron pocket at $(\pi,0)$ only over a very small range of angles on the
$(\pi,0)$ pocket. Instead, the influence of the superconducting gap is to shift and sharpen the normal state peak such that spectral weight at low energies is removed and effectively a gapped paramagnon appears. The final peak position of the apparent ``neutron resonance'' as shown in Fig. \ref{fig:nresonance} is roughly given by the sum of the gaps on the pockets at $\Gamma$ and $(\pi,0)$.

 \section{Penetration depth anisotropy}
 \begin{figure}[tb]
 \includegraphics[width=\columnwidth]{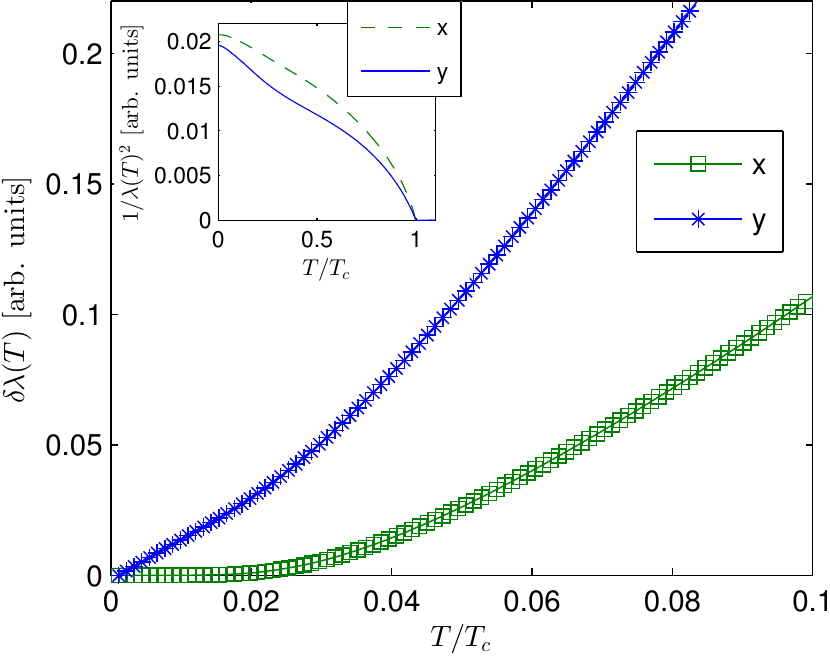}
\caption{(Color online)
Penetration depth $\delta\lambda(T)=\lambda(T)-\lambda(0)$ at low $T$ showing power-law (linear) $T$ dependence in the $x$ ($y$) direction where the nodes are not (are) captured. Inset: Corresponding components of the superfluid density tensor $\propto 1/\lambda(T)^2$ calculated over a wide range of $T$ assuming a mean-field-like superconducting order parameter as a function of $T$.
\label{fig:pen}}
\end{figure}
A possible experimental probe that could confirm and elaborate on
the nodal nature of the superconducting gap measured by STM is the London penetration depth. Since the system under discussion shows a strong deviation from a $C_4$-symmetric electronic structure, this deviation should also imprint itself on the superconducting order parameter, yielding a gap that is a combination of $s$- and $d$-wave contributions as obtained in our calculation. Indeed, the penetration depth $\lambda$ should display an anisotropy that is directly related to the superconducting order parameter. To make predictions for our proposed gap state, we calculate the penetration depth from the current-current correlator as follows\cite{Sheehy04,Eremin10}:
Assuming that contributions from individual bands $\nu$ at the Fermi level simply sum up in the correlator\cite{Mishra11}, we obtain for the current direction $i$
 \begin{equation}
  \frac 1{\lambda_i^2}=\sum_\nu \frac 1 {\lambda_{\nu,i}^2}\;.
 \end{equation}
The corresponding result from a single band calculation is given by
 \begin{align}
  \frac 1{\lambda_{\nu,i}^2}&=\frac{4\pi e^2}{c^2\hbar^2}\sum_{\bf{k}}\frac{d\xi_\nu(\bf{k})}{dk_i}\biggl(\frac{d\xi_\nu({\bf{k}})}{dk_i} |\Delta_{\bf{k}}|^2-\frac{d|\Delta_{\bf{k}}|}{dk_i}|\Delta_{\bf{k}}|\xi_\nu(\bf{k})\biggr)\notag\\
  &\times\frac 1{E_{\nu,\bf{k}}^2}\biggl(\frac 1{E_{\nu,\bf{k}}}\tanh\bigl(\frac{E_{\nu,\bf{k}}}{2 k_B T}\bigr) -\frac{1}{2k_BT}\mathop{\text{sech}^2}\bigl(\frac{E_{\nu,\bf{k}}}{2 k_B T}\bigr)\biggr).
  \label{eq_pen}
 \end{align}
Evaluating Eq.~(\ref{eq_pen}) for $\approx 10^6$ $k$ points as in the calculation of the susceptibility in the superconducting state, we set $T_c=9\;\text{K}$ and use a
mean-field-like $T$ dependence of the superconducting order parameter, setting $\Delta_{\bf{k}}=g({\bf k})\Delta_0\tanh(1.76\cdot\sqrt{T_c/T - 1})$.\cite{Eremin10}
In the inset of Fig.~\ref{fig:pen} we show the result for the two eigenvalues of the superfluid density tensor, $\propto 1/\lambda(T)^2$, corresponding
to current in the $x$ and $y$ directions, respectively.
For the choice $i=y$ the nodal region can be probed with finite ${d\xi_\nu({\bf{k}})}/{dk_y}$ yielding a linear penetration depth at low $T$. On the other hand, the absence of nodes in the $x$ direction is reflected in the missing linear term in the penetration depth such that the finite but small gap yields a power law similar to $T^3$ as expected when the direction of the current is perpendicular to the nodes\cite{Gross86}. Note that the crossover energy scale is set not by the maximum of the gap, but by the smaller minimal gap on the $\Gamma$-centered Fermi surface. As  discussed previously in Ref.~\onlinecite{Mishra11} (and confirmed in our calculations), the gap velocity ${d|\Delta_{\bf{k}}|}/{dk_i}$ can be safely neglected in the calculation.

 \section{Discussion}

Our model, based on fits to ARPES and QO data, provides an explanation of the observed splitting of the Knight shift and $(T_1T)^{-1}$ as seen in NMR experiments, is consistent with the $T$ and $\omega$ evolution of the INS data, and predicts a superconducting order parameter within the spin-fluctuation pairing scenario consistent with STM.    The striking result is that the $T$ evolution  of all  these properties  can be explained by introduction of a simple, phenomenological  orbital order with mean-field like temperature dependence.\cite{MKHA_PRL15,Note1}  Of course we have not specified the physical origin of this orbital ordering here, but it is interesting to note that the system found to fit experiments appears to be very close to two competing magnetic instabilities, at $(\pi,0)$, and $(\pi,q)$, consistent with the proposal of Glasbrenner {\it et al}.\cite{Glasbrenner15} that long-range magnetism is suppressed in this system by closely competing magnetic states. If this is true, the orbital ordering effect could indeed be driven by spin fluctuations in
the high-$T$ tetragonal state.  We remind the reader, however, that rather small changes in band structure led to a different magnetic response in our earlier work, so for the moment this conclusion should be treated with caution.

All current microscopic theories of orbital ordering rely on an interplay of orbital fluctuations and spin fluctuations, and generally link the structural and magnetic transitions. Recently, a theory of the nematic transition in FeSe without long-range magnetic
order was proposed in Refs. \onlinecite{Yamakawa2015},\onlinecite{Onari2015}, but it is not clear whether it yields electronic structure at low $T$ consistent with ARPES and QO data.

We now ask why orbital ordering favors low-energy spin fluctuations at $(\pi,0)$.
In our model, it appears that differences in the magnetic responses in the $x$ and $y$ directions below $T_S$ due to the lowered symmetry drive the system to a state where a $(\pi,0)$ instability competes very closely with a state at $(\pi,q)$. In Ref. \onlinecite{Glasbrenner15} it is argued that this competition suppresses the magnetic ordering temperature. The transfer of spectral weight from $(0,\pi)$ to $(\pi,0)$ with onset of orbital order as seen in Fig.~\ref{fig:staticChi} is one of the consequences of increased nesting, e.g., that $\xi_\nu(\bf{k})=-\xi_\mu(\bf{k}+\bf{q})$ is fulfilled for a larger set of $k$ points. Unlike in a single-band picture where Fermi surface nesting (at zero energy) automatically gives rise to a logarithmic singularity in the Lindhard function, however, in multiband systems significant contributions
to the susceptibility can arise from finite energy scattering. In general, this complicates the identification of dominant nesting vectors for the multiband case, and can lead to a situation where the susceptibility at two different momenta exhibits an opposite trend than expected from the Fermi surface. Looking for example at Fig. \ref{fig:bandfscut}(a) a better Fermi surface nesting at ${\bf q}_2=(0,\pi)$ than at ${\bf q}_1=(\pi,0)$ can be seen, but the susceptibility is larger at ${\bf q}_1$.
  \begin{figure}[tb]
\includegraphics[width=\columnwidth]{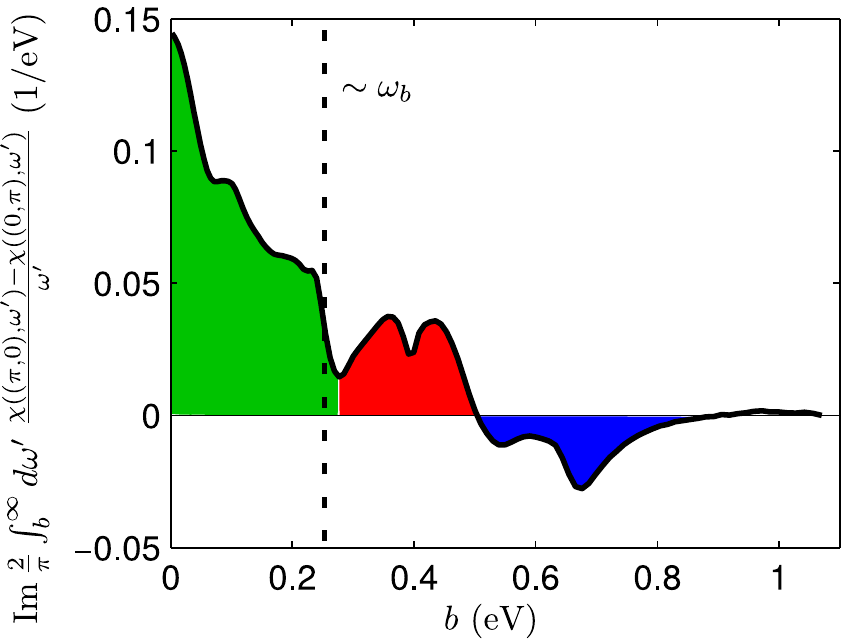}
\caption{(Color online)
Analysis of the redistribution of weight in the real part of the static susceptibility using the Kramers-Kronig relation in Eq.~(\ref{eq_kk}): The partial integral down to energies of $\omega_b\approx 200\,\text{meV}$ does not accumulate large differences (red and blue areas), but at low energies there is a significant contribution that makes the susceptibility at $(\pi,0)$ larger than at $(0,\pi)$.
\label{fig_kk}}
\end{figure}

One way to determine the origin of the larger weight at ${\bf q}_1$ is to 
use the Kramers-Kronig relation that connects the real and imaginary parts of the susceptibility
\begin{equation}
  \mathop{\text{Re}} \chi({\bf q},0)=\frac 2 \pi \int_0^\infty d\omega\;\frac{\mathop{\text{Im}} \chi({\bf q},\omega)}{\omega}\,,
  \label{eq_kk}
 \end{equation}
and identify the relevant energies where the integrand on the right-hand-side largely contributes. One key question is the origin of the splitting of the $(\pi,0)$ and $(0,\pi)$ response. A plot of the difference of the integrand in Eq.~(\ref{eq_kk}) as a function of the lower boundary $b$ reveals that high-energy fluctuations are not affected by the orbital order, but scattering at an energy transfer below $\omega_b\approx 200 \,\text{meV}$ gives rise to a larger susceptibility at $(\pi,0)$ as shown in Fig.~\ref{fig_kk}. The major contributions are made up by scattering of states dominantly of $d_{yz}$ orbital character with momentum transfer $(\pi,0)$ which is significantly larger than the corresponding scattering of the states of $d_{xz}$ orbital character with momentum transfer $(0,\pi)$.

As expected from the electronic structure, the superconducting gap does not display tetragonal symmetry and is a linear combination of $s$- and $d$-wave symmetries allowing for accidental nodes. The calculated state shown in Fig.~\ref{fig_SC} indeed has vertical line nodes on the X-centered electron pocket where the gap reaches zero such that the DOS becomes V-shaped at low energies; see Fig. \ref{fig_SC}(b). We note that the gap function $g({\bf k})$ changes sign on the extremely 2D X-centered electron pocket, but only over a very small region in momentum space as it can be seen from the areas in Fig.~\ref{fig_SC}(a) marked in black where the gapfunction is very small. In the absence of a sign change in the gap function $g({\bf k})$ for momentum transfer $(\pi,0)$, we argue that the corresponding susceptibility should not show a strong resonance of the usual $s_\pm$ type, but rather a peak due to the proximity to the magnetic instability in the normal state which is then modified in the superconducting state by gapping out states at the Fermi level; see Fig.~\ref{fig:nresonance}.

The gap state with accidental nodal lines can be probed by the anisotropy of the penetration depth. Unlike symmetry-imposed nodes as, for example, in the $d$-wave state of cuprate superconductors, the nodes only occur in the $y$ direction, but not in the $x$ direction. This is a simple consequence of Eq.~(\ref{eq_pen}) where, for $i=x$, the derivative $d\xi_\nu({\bf{k}})/dk_x$ at a nodal line vanishes, and all the other terms in the $k$ sum stem from fully gapped quasiparticle states. The authors of Ref.~\onlinecite{Kang14} have studied the distortion from the tetragonal state by strain and resulting movements of nodal points in the gap structure and proposed observable $T\rightarrow 0$ penetration depth anisotropies in consequence.  In our case, the distortion of the electronic structure is intrinsic, and we emphasize  that also the $T$ dependence of the penetration depth, which should be easier to observe, exhibits qualitative anisotropy arising from the unusual nodal structure of the gap function.

\section{Conclusions}
FeSe is one of the most challenging and intriguing of the Fe-based superconductors. In this work we began with a
tight-binding model which provided a good fit to both ARPES and QO data on this system both at high and low $T$, but assuming a phenomenological orbital ordering. However the dynamical magnetic susceptibility derived from this model proved inconsistent with subsequent INS measurements, so we posed the question of whether small local adjustments to the tight-binding Hamiltonian could preserve the good fit to one-particle properties while simultaneously fitting
two-particle susceptibility within an RPA approach. Our slightly modified model provided an equally good fit to ARPES and QO data and explains many aspects of the INS measurements semiquantititatively such as the transfer of spectral weight from N\'eel to stripe fluctuations and a paramagnon dispersion along $(\pi,0)$ to $(\pi,\pi)$. While our approach does not constitute a complete microscopic theory, we  have provided an internally self-consistent model of the electronic structure and spin excitations which evolves with $T$ according to experiment. This has the principal virtue of  confirming that the main physical phenomenon driving the subtle evolution of the electronic excitations is orbital ordering.
We have proposed that both site-centered and bond-centered orbital order of similar order of magnitude are required to explain the existing experimental data. In the superconducting state, we find an order parameter with accidental nodes leading to a V-shaped DOS and an anisotropy in the penetration depth as well as an enhanced magnetic response at momentum transfer $(\pi,0)$.

\section{Acknowledgements}

We acknowledge our useful discussions with A. Boothroyd, J. Zhao, B. B\"uchner, M. H. Christensen, \mbox{A. T. R{\o}mer}, and H. Kontani. A. K. and B. M. A. acknowledge financial support from a Lundbeckfond fellowship (Grant No. A9318). P. J. H.  was partially supported by the Department of Energy under Grant No. DE-FG02-05ER46236.

\appendix
\section{Hamiltonian: Band structure, Interactions}
\begin{figure}[t]
\includegraphics[width=\columnwidth]{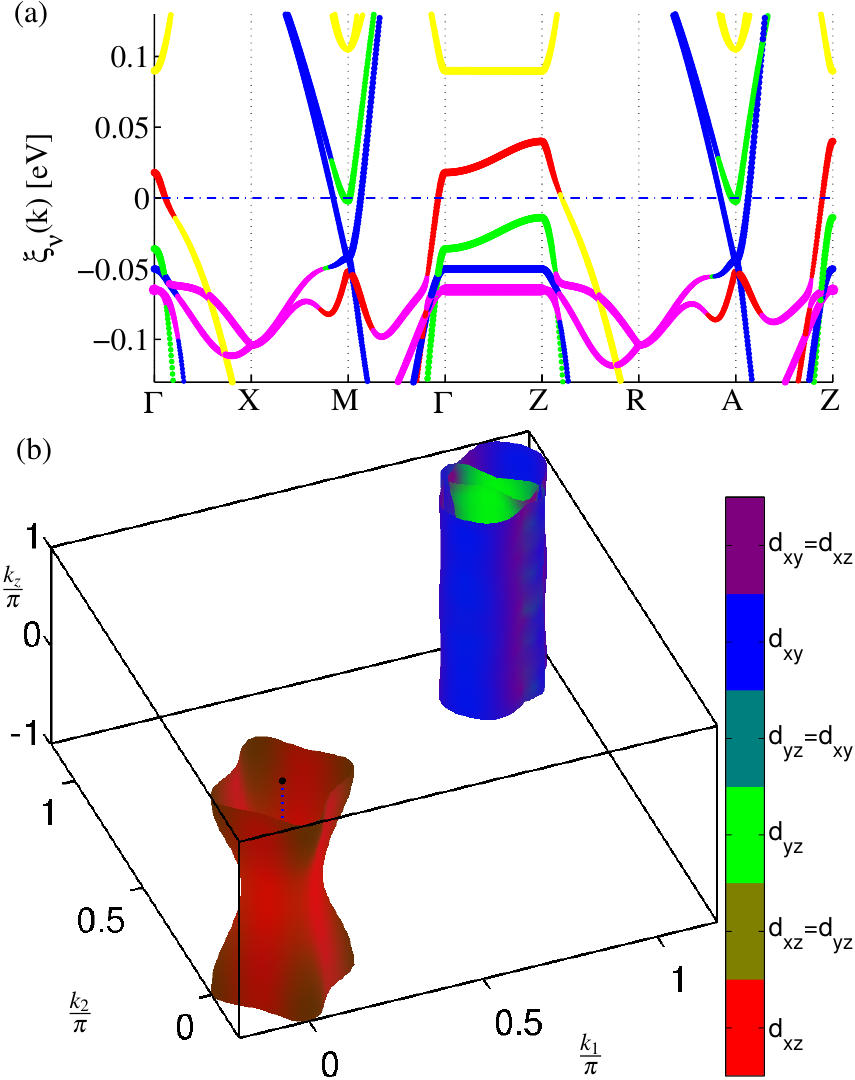}
\caption{(Color online) (a) Low energy band structure of FeSe including a spin-orbit coupling of approximately $20\,\text{meV}$ at $T=40\,\text{K}$ showing the band splitting due to orbital ordering where the main orbital character is indicated by the colors red $d_{xz}$, green $d_{yz}$, blue $d_{xy}$, yellow $d_{x^2-y^2}$, and purple  $d_{3z^2-r^2}$, and the thickness represents the magnitude of the main orbital character. (b) Fermi surface for the same bands in the folded Brillouin zone that yields the frequencies for quantum oscillations as shown in Fig. \ref{fig:QO}.}
\label{fig:bandfs}
\end{figure}
\begin{figure}[t]
\includegraphics[width=\columnwidth]{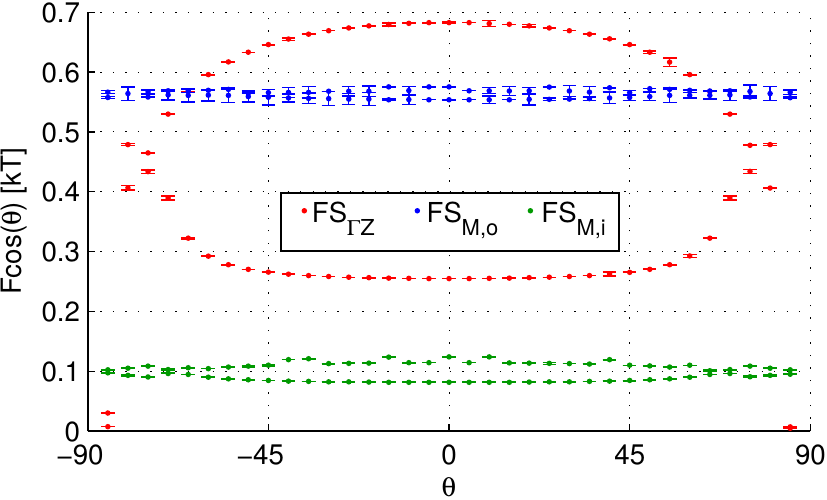}
\caption{(Color online) QO frequencies as a function of magnetic field angle $\theta$ for our band structure at low $T$ including spin-orbit coupling obtained using a numerical method\cite{MKHA_PRL15, Rourke12, Diehl14,Backes14} where the error bars indicate the numerical uncertainty in the determination of the extremal orbits.}
\label{fig:QO}
\end{figure}
\begin{figure}[t]
\includegraphics[width=\columnwidth]{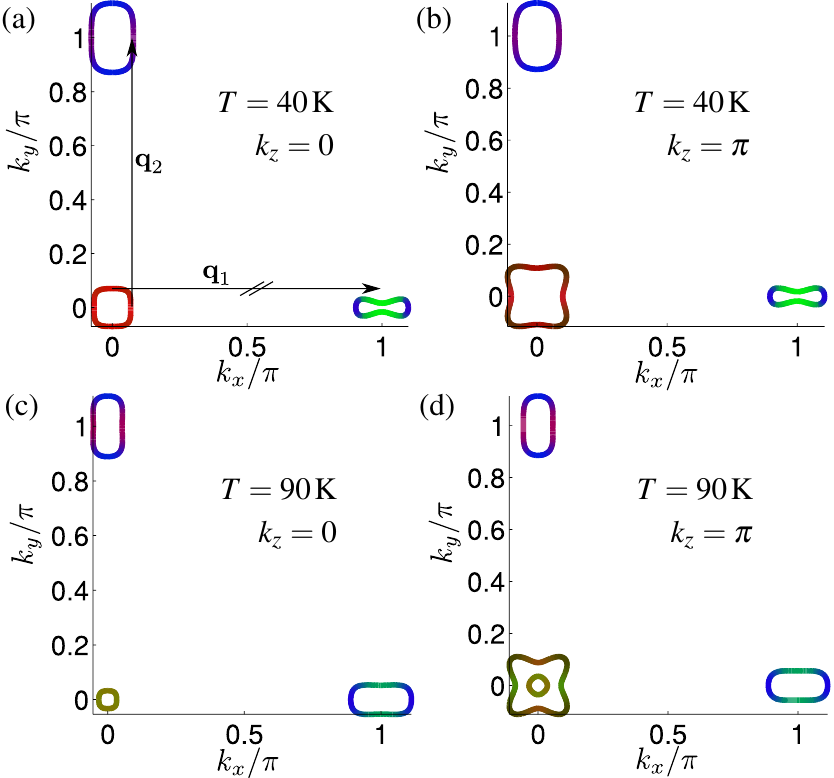}
\caption{(Color online) Cuts of the Fermi surface at $k_z=0$ (left) and $k_z=\pi$ (right) for $T=40\,\text{K}$ (top) and  $T=90\,\text{K}$ (bottom) in the presence of spin-orbit coupling. Color code as in Fig.~\ref{fig:bandfs}.
In (a) it is visualized that the Fermi surface nesting for momentum transfer ${\bf q}_2$ is larger than for  ${\bf q}_1$.}
\label{fig:bandfscut}
\end{figure}

The band structure parameters utilized in the current paper are very similar to those presented in the Supplemental Material of Ref. \onlinecite{MKHA_PRL15}. In the present study we have included additional shifts $\Delta t^{11}_{34}=0.066, \Delta t^{10}_{23}=-0.023i,$ and $\Delta t^{10}_{13}=-0.014i$ since this leads to a better agreement with recent neutron scattering experiments\cite{Boothroyd15,Bourges15}. The momenta $(k_x,k_y)$ and $(q_x,q_y)$ are measured in units of the inverse lattice constants along the two orthorhombic crystal directions and refer to the Brillouin zone corresponding to the iron lattice, except for Fig. \ref{fig:bandfs} where the 2-Fe Brillouin zone has been used. The properties in terms of Fermi surface and quantum oscillation frequencies are mostly unchanged compared to those presented earlier; see Figs.~\ref{fig:bandfs} and \ref{fig:QO}. The main effect of
these quite small changes is to move the leading magnetic instability from $(\pi,\pi)$ in Ref. \onlinecite{MKHA_PRL15} to $(\pi,0)$ in the present case. Note that it is necessary to include a spin-orbit coupling for capturing a band splitting such that only one hole-like pocket is present at the $\Gamma$ point at high $T$\cite{fernandes14}; see Fig. \ref{fig:bandfscut}. For the low-$T$ electronic structure where the bands are split by the orbital order, the spin-orbit coupling has little effect and is neglected for the calculation of physical quantities at low $T$ for simplicity.

The local interactions are included via the Hubbard-Hund Hamiltonian
\begin{align}
	H_U =& {U}\sum_{i,\ell}n_{i\ell\uparrow}n_{i\ell\downarrow}+{U}'\sum_{i,\ell'<\ell}n_{i\ell}n_{i\ell'}
	\nonumber\\
	& +  {J}\sum_{i,\ell'<\ell}\sum_{\sigma,\sigma'}c_{i\ell\sigma}^{\dagger}c_{i\ell'\sigma'}^{\dagger}c_{i\ell\sigma'}c_{i\ell'\sigma}\\
	& +  {J}'\sum_{i,\ell'\neq\ell}c_{i\ell\uparrow}^{\dagger}c_{i\ell\downarrow}^{\dagger}c_{i\ell'\downarrow}c_{i\ell'\uparrow} \nonumber,
\end{align}
where we set $J'=J$, $U'=U-2J$ to restrict ourselves to the spin-rotation-invariant case.

\section{Nuclear magnetic resonance}
\begin{figure}[b]
\includegraphics[width=\columnwidth]{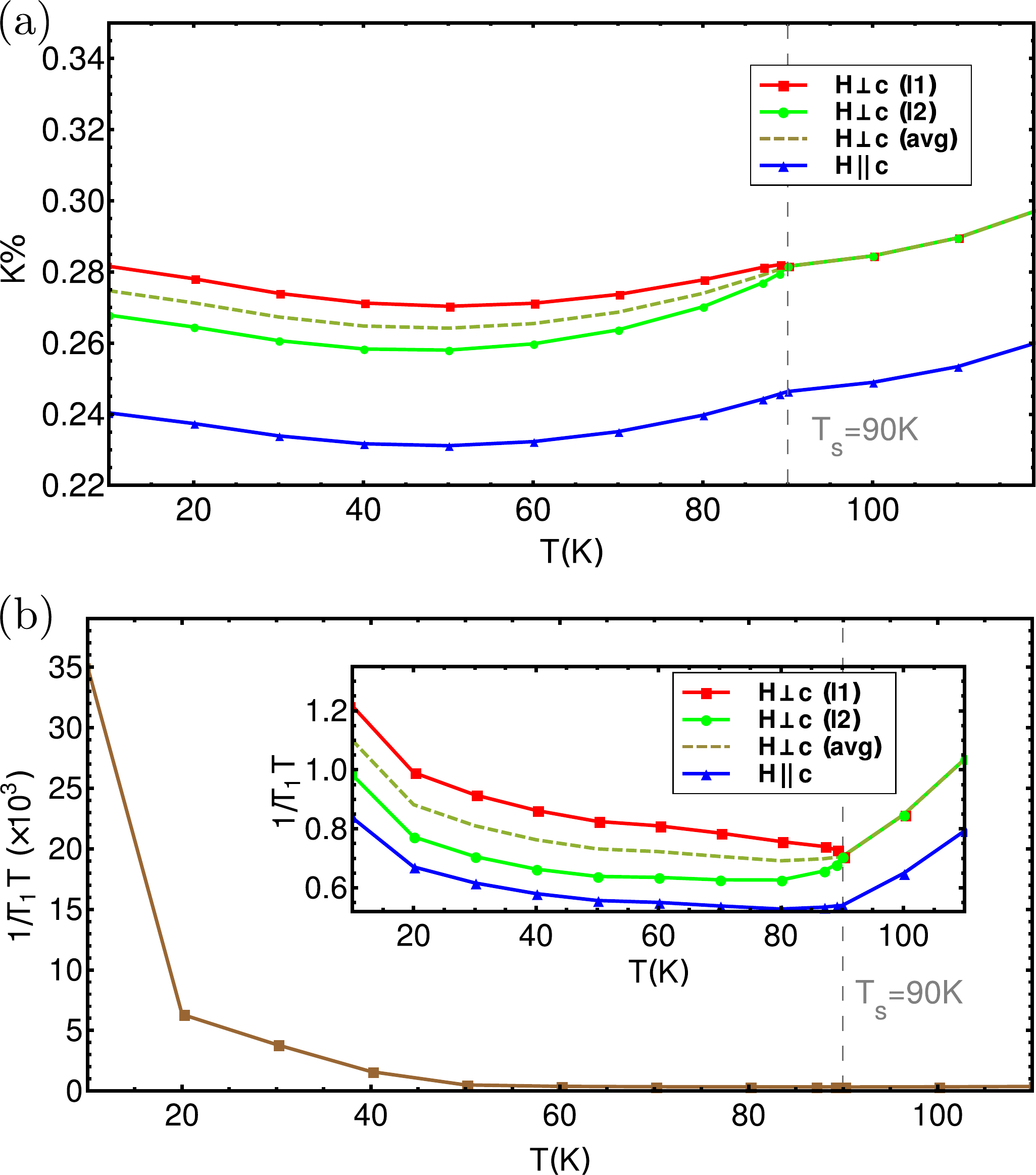}
\caption{(Color online) (a) $T$ dependence of the Knight shift showing the splitting of the Knight shift signal below $T_S$. Here $l1$ and $l2$ are the two orthorhombic domains and the $c-$axis is perpendicular to the FeSe plane. (b) Spin lattice relaxation rate $1/T_1T$ calculated without the structure factor showing the strong upturn below $\sim 50\,\text{K}$.
Inset: $1/T_1T$ calculated with the structure factor screening calculated with the correct expression for $A^{\xi\psi}_{hf}({\bf{q}})$ in Eq. (\ref{eq_nmr}) showing the effects of twin orthorhombic domains.}
\label{fig:nmr}
\end{figure}
NMR experiments in FeSe have probed the $^{77}$Se atom and mapped out both the $T$ dependence of the Knight shift and the low-energy spin fluctuations through measurements of the spin lattice relaxation rate ($1/T_1T$)\cite{baek14,bohmer15}. Both the Knight shift and $1/T_1T$ are split below the structural transition due to contributions from the two orthorhombic domains in twinned FeSe samples. It is found that spin fluctuations measured by $1/T_1T$ are not affected by the structural transition itself but are enhanced at much lower $T$.
We have calculated the NMR Knight shift
as $A_{hf}\chi_{\text{RPA}}(\bf{q}=0)$ with a constant hyperfine form factor $A_{hf}$ above the structural transition and an order-parameter-like $T$ dependence in the orbitally ordered state\cite{baek14,MKHA_PRL15}. As seen in Fig~\ref{fig:nmr}(a) by fitting the magnitude of the form factor we get a good agreement with Knight shift measurements similar to the results presented in Ref.~\onlinecite{MKHA_PRL15}.
The $T$ dependence of the spin lattice relaxation rate $1/T_1T$ can be evaluated from the expression,
\begin{align}\frac{1}{T_1T}=\lim_{\omega_0\rightarrow 0}\frac{\gamma_N^2}{2N}k_B\sum_{{\bf{q}} \xi \psi} |A^{\xi\psi}_{hf}({\bf{q}})|^2
\frac{\mathop\text{Im} \{\chi^{\xi \psi}_{\text{RPA}}({\bf{q}},\omega_0)\}}{ \hbar \omega_0}.
\label{eq_nmr}
\end{align}
Here $A^{\xi\psi}_{hf}({\bf{q}})$ is the $q$-dependent form factor, $\xi,\psi$ indices run over the Cartesian
coordinates. In the paramagnetic state the form factor matrix is diagonal implying that $\xi=\psi$.  In the orthorhombic phase, we have $A^{xx}_{hf}({\bf{q}})=A^{yy}_{hf}({\bf{q}})\neq A^{zz}_{hf}({\bf{q}})$.   Assuming a nearest neighbor interaction between the nuclear spin of the $^{77}$Se atom and the iron conduction electrons, we express the form factor as $A^{\psi\psi}_{hf}({\bf{q}})=(\alpha^{\psi\psi}(T)+\beta^{\psi\psi}(T))\cos(q_x/2+q_y/2)+(\alpha^{\psi\psi}(T)-\beta^{\psi\psi}(T))\cos(q_x/2-q_y/2)$.  We obtain a
good agreement with both the Knight shift and $1/T_1T$ experiments for $\alpha^{\psi\psi}(T)=m^{\psi\psi}[1+g(t)]$, $\beta^{\psi\psi}(T)=m^{\psi\psi}g(t)$, where $m^{xx}=m^{yy}=0.4$, $m^{zz}=0.35$, and $g(t)$ is a mean-field $T$ dependence which is zero above the structural transition\cite{Note1}. As can be seen from Fig.~\ref{fig:nmr}(b) the spin lattice relaxation rate is enhanced at low $T$ compared to the structural transition.

%

\end{document}